\documentclass[aps,prl,twocolumn,10pt,showpacs]{revtex4-1}

\usepackage{amsmath}
\usepackage{graphicx}
\usepackage{times}
\usepackage{amssymb}
\usepackage{hyperref}
\usepackage{textcomp}
\usepackage{xcolor}
\usepackage{pdfpages}

\begin{document}
\title{Optimal Squeezing in Resonance Fluorescence via Atomic-State Purification}
\author{P. Gr\"unwald}
\email{Electronic address: peter.gruenwald2@uni-rostock.de}
\author{W. Vogel}
\affiliation{Arbeitsgruppe Quantenoptik, Institut f\"ur Physik, Universit\"at Rostock, D-18055 Rostock, Germany}
\date{\today}

\begin{abstract}
Squeezing of atomic resonance fluorescence is shown to be optimized by a properly designed environment, 
which can be realized by a quasi-resonant cavity.
Optimal squeezing is achieved if the atomic coherence is maximized, corresponding to a pure atomic quantum state. The atomic-state purification 
is achieved by the backaction of the cavity field on the atom, which increases the atomic coherence and decreases the atomic excitation.
For realistic cavities, the coupling of the atom to the cavity field yields a purity of the atomic state of more than 99\%. The fragility of squeezing against dephasing is substantially reduced in this scenario, which may be important for various applications.
\end{abstract}

\pacs{42.50.Pq, 37.30.+i, 42.50.Ct}

\maketitle

\paragraph*{Introduction.}
A single atom and its coupling to the electromagnetic field is a system 
of fundamental interest for the understanding of the quantum phenomena of light and matter.
It was predicted that a driven two-level atom emits antibunched light~\cite{carm}, 
which cannot be described by the classical Maxwell theory.
The first experimental demonstration of this effect was based on the resonance fluorescence of an atomic beam~\cite{ki-dag-ma}. In a related experiment it was demonstrated that a sub-Poissonian photon statistics may occur~\cite{short}.
Later on, photon antibunching 
could be demonstrated with single trapped ions~\cite{walther,toschek}.

Squeezing was predicted to occur in the single-atom resonance fluorescence~\cite{walls}. It can also be realized in the fluorescence of many atoms, via regular arrangement of the atoms~\cite{vowe}, detection in the forward direction with respect to the pump-beam~\cite{heire}, and  bistability in a strong driving field~\cite{Reid}. The latter two cases, could be experimentally demonstrated~\cite{LuBali,Raizen}. Squeezing in single-atom resonance fluorescence could not be observed yet. Based on homodyne correlation measurements with a weak local oscillator, an efficient measurement technique was proposed~\cite{vogel}, which is not limited by the collection efficiency of the fluorescence light. Its feasibility was demonstrated in resonance fluorescence experiments~\cite{Blatt}.

Very recently, squeezed light has been observed in the output channel of a weakly driven high-Q cavity, containing a single atom~\cite{Rempe}. Since an empty driven cavity cannot produce squeezing, the observed squeezing of the cavity output field is clearly based on the coupling of the atom to the cavity field. This work demonstrates, that squeezed light originated from a single atom is a prevailing subject. The squeezed light in the output field of a cavity containing a single atom is undoubtedly a fundamental issue for its own. Under general conditions, however, one cannot expect that this is equivalent to the demonstration of squeezing in resonance fluorescence. Hence a direct fluorescence measurement would still be of
fundamental interest.
 
The present Letter deals with the optimization of atomic resonance fluorescence with respect to squeezing. 
It is shown that squeezing becomes optimal, when the atomic coherence is perfectly controlled, which is equivalent to the purification of the atomic quantum state. 
Atomic coherence control and state-purification can be realized by a properly adjusted cavity, even for different choices of the mean atomic excitation. This implies the possibility to control the fluorescence intensity to some extend. Similar to ordinary fluorescence measurements, the light is observed out the side of the cavity. The setup under study substantially reduces the fragility of squeezing against dephasing.
This achievement is expected to be of great relevance for applications in miniaturized systems, such as quantum dots in semiconductor microcavities.

\paragraph*{Optimal squeezing.} Let us consider a general two-level atom (TLA) in an arbitrary environment. The atomic source field can be written as
\begin{equation}
  \hat E_{\rm s}=|\chi|(\hat A_{12}e^{i\phi}+\hat A_{21} e^{-i\phi}),
\end{equation}
where $\hat A_{ij}=|i\rangle\langle j|$ ($i,j=1,2$) is the flip operator
of the atom with ground state $|1\rangle$ and excited state $|2\rangle$. 
Here $|\chi|$ describes the atom-light-field coupling and the phase $\phi$ includes the phase of the driving laser.

The field is squeezed, if the normally ordered field variance becomes negative, corresponding to a noise reduction below the vacuum level. Optimizing with respect to the phase, we get for a TLA
\begin{align}
  \frac{\langle:(\Delta\hat E)^2:\rangle}{|\chi|^2}=&2(\langle\hat A_{22}\rangle-2|\langle\hat A_{12}\rangle|^2)\label{eq.Atvarsol1},
\end{align}
see~\cite{WelVo}.
The full field,
$\hat E=\hat  E_{\rm f}+\hat  E_{\rm s}$, is composed of the free field $\hat  E_{\rm f}$ (assumed to be in the vacuum state) and the atomic source field $\hat  E_{\rm s}$. The ``$:\cdots:$'' prescription denotes normal ordering. 

The atomic expectation values in Eq.~(\ref{eq.Atvarsol1}) are readily derived from the atomic density operator,
$\hat \sigma = \sum_{ij} \sigma_{ij} \hat A_{ij}$. The density matrix $\sigma$ reads as
\begin{equation}
  \sigma=\left(\begin{array}{cc}
		    \langle\hat A_{11}\rangle & \langle\hat A_{21}\rangle\\ \langle\hat A_{12}\rangle & \langle\hat A_{22}\rangle
                   \end{array}
\right).
\end{equation}
From the positive semidefiniteness of quantum states it follows that $\det \sigma \ge 0$, that is
\begin{equation}
  |\langle\hat A_{12}\rangle|^2\leq\langle\hat A_{11}\rangle\langle\hat A_{22}\rangle=\langle\hat A_{22}\rangle-\langle\hat A_{22}\rangle^2.\label{eq.CSI}
\end{equation} 
Here we made use of the completeness relation of the TLA, $\sum_i \hat A_{ii} =\hat 1$.
This inequality defines the maximal atomic coherence, $|\langle\hat A_{12}\rangle|^2$,
for any atomic excitation $\langle\hat A_{22}\rangle$.

Using this result, for arbitrary atomic excitation the minimal variance follows for the maximal atomic coherence
as
\begin{equation}
\frac{\langle:(\Delta\hat E)^2:\rangle_{\text{min}}}{|\chi|^2}=2\langle\hat A_{22}\rangle(2\langle\hat A_{22}\rangle-1).\label{var-min}
\end{equation} 
The absolute minimum follows for $\langle\hat A_{22}\rangle=1/4$,
\begin{equation}
  \frac{\langle:(\Delta\hat E)^2:\rangle_{\text{abs}}}{|\chi|^2}=-\frac{1}{4}. \label{abs}
\end{equation}
This value cannot be attained in free-space resonance fluorescence~\cite{walls,WelVo}.

Let us consider the structure of the atomic quantum state for optimal squeezing.
The purity of the state is given by
\begin{equation}
  \text{Tr}\{\hat \sigma^2\}=1-2(\langle\hat A_{22}\rangle-\langle\hat A_{22}\rangle^2-|\langle\hat A_{12}\rangle|^2).\label{eq.purity}
\end{equation}
Compared with Eq.~(\ref{eq.CSI}), purity of the atomic state is equivalent to 
maximal atomic coherence. That is, optimal squeezing of the resonance fluorescence is achieved for a pure state of the atomic subsystem. Hence, to optimize squeezed emission from a single atom, the task is to find an environment for which the atomic subsystem is in a (nearly) pure state.

Note that the above results may be used to estimate the squeezing in the resonance fluorescence of a TLA, without the need of homodyne detection. The minimal variance, optimized with respect to the phase,
\begin{equation}
	\frac{\langle:(\Delta\hat E)^2:\rangle}{|\chi|^2}=\frac{\langle:(\Delta\hat E)^2:\rangle_{\text{min}}}{|\chi|^2}+2(1-\text{Tr}\{\hat\sigma^2\}),\label{eq.Sq-Pur}
\end{equation}
can be expressed by the minimal variance for maximal atomic coherence (first term),
and the atomic-state purity (occurring in the second term). The first contribution is solely determined by the atomic excitation, cf. Eq.~(\ref{var-min}).
The excitation can be observed by comparison with the saturation intensity. Simple methods also exist for detecting the purity of low-dimensional systems~\cite{Filip,Ekert,Nakaz}. In view of the relevance of purity measurements in the field of Quantum Information, this may open an alternative possibility to infer squeezing in resonance fluorescence from independent measurements.

\begin{figure}[h]
  \includegraphics[width=7cm]{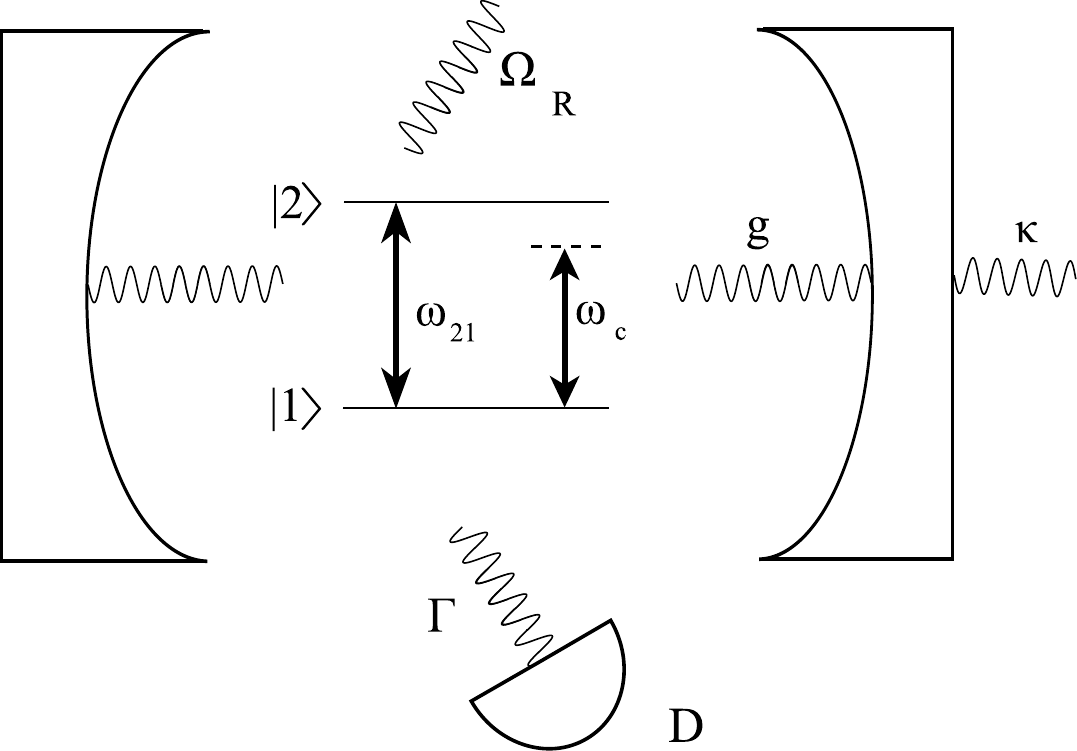}
  \caption{Sketch of the coherently driven TLA inside a quasi-resonant cavity with losses. The fluorescent light is detected (D) out the side of the cavity. Wavy lines indicate light fields driving the atom, emitted into the cavity, out of the cavity, or out the side of the cavity. Straight arrows indicate the frequencies of the atomic resonance and the quasi-resonant cavity mode.}\label{fig.sys}
\end{figure}

\paragraph*{Squeezing in cavity-assisted fluorescence.} In the following we will deal with a method for the practical realization of optimal squeezing by atomic-state purification.
We will show that a single-mode cavity may serve as an environment of the TLA to optimize squeezing in resonance fluorescence. In fact, our system may purify the atomic state for different values of the atomic excitation $\langle\hat A_{22}\rangle$. The cavity-assisted optimization of squeezing from a TLA is feasible with current technology, cf. the required scheme in Fig.~\ref{fig.sys}.

Consider the TLA being driven by a coherent light source of frequency $\omega_\text L$, with the Rabi frequency $\Omega_\text R$ (chosen to be real). Its spontaneous emission is characterized by the energy relaxation rate $\Gamma$. Furthermore the atom is coupled to a single mode cavity of frequency $\omega_\text c$, with coupling strength $g$. The cavity excitation is described by bosonic creation and annihilation operators, $\hat a^\dagger$ and $\hat a$, respectively. The cavity emits light with a rate $\kappa$. The Hamiltonian for this system, in the frame rotating with $\omega_\text L$ and in the rotating-wave approximation, reads as
\begin{equation}
  \hat H/\hbar=\delta_\text a\hat A_{22}+\delta_\text c\hat a^\dagger\hat a+g(\hat a^\dagger\hat A_{12}+\hat A_{21}\hat a)+\Omega_\text R(\hat A_{12}+\hat A_{21}),\label{eq.Hamilton}
\end{equation}
where $\delta_\text a=\omega_{21}-\omega_\text L$ and $\delta_\text c=\omega_\text c-\omega_\text L$. The density operator $\hat\varrho$ of the full system obeys the von-Neumann equation with Lindblad terms for the different decay channels,
\begin{align}\label{eqmo}
    \frac{d\hat\varrho}{dt}&=\frac{1}{i\hbar}[\hat H,\hat\varrho]+\frac{\Gamma}{2}\mathcal{L}_{\hat A_{12}}[\hat\varrho]+\frac{\kappa}{2}\mathcal{L}_{\hat a}[\hat\varrho],\\
    \mathcal{L}_{\hat X}[\hat\varrho]&=2\hat X\hat\varrho\hat X^\dagger-\hat X^\dagger\hat X\hat\varrho-\hat\varrho\hat X^\dagger\hat X.
\end{align}
In general this system cannot be solved analytically.  The numerical calculations are based on solving the steady-state equations of the density matrix, by truncation at a sufficiently large number of cavity photons, for details see~\cite{Supp}.

Let us now consider the cavity-assisted scenario for a strong driving field, $\Omega_\text R\gg g,\Gamma,\kappa$, but detuning of the atom by $\delta_\text a\approx\Omega_\text R$. Hence the effective pumping is still limited, so that no saturation occurs, $\langle\hat A_{22}\rangle<1/2$. A similar scenario has been studied in~\cite{FrQu,QuFr}, but not in the context of squeezing.
Under such conditions the Mollow-triplet is clearly visible in the spectrum, and the sidebands are well separated, at (nearly) the same frequencies as in free space:
\begin{equation}
  \omega_0=\omega_\text L,\quad\omega_{\pm}=\omega_\text L\pm\sqrt{(2\Omega_\text R)^2+\delta_\text a^2}.\label{eq.RFsidebands}
\end{equation}
Following the argumentation in~\cite{FrQu}, for a detuning of the cavity mode of $\delta_\text c\approx\delta_\text a$ its excitation depends on the ratio of the atomic decay to the cavity damping. If the cavity emission rate significantly exceeds the atomic decay, the excitation of the cavity is proportional to $\Gamma/\kappa$, hence the cavity is almost empty. In this case the outcoupling of the cavity photons is fast compared to the emission rate of the atomic fluorescence out the side of the cavity. It should be noted that we do not need a very good cavity, solely one with $\kappa\gg\Gamma$.

Despite the tiny excitation, the cavity effects become visible if a fluorescence sideband is close to a cavity resonance. Single-photon transitions in the cavity occur if the detuning, $\delta_\text c$, is resonant to a sideband according to Eq.~(\ref{eq.RFsidebands}), 
\begin{equation}
  \delta_\text c^2=(2\Omega_\text R)^2+\delta_\text a^2.\label{eq.cavres}
\end{equation}
At such a resonance, besides the fluorescence out of the side of the cavity, we also obtain enhanced emission of the cavity itself. The excitation of the cavity increases slightly, which is consistent with the argumentation in~\cite{FrQu}.  Nevertheless, the cavity emission increases strongly, as it scales with $\kappa$ which is large compared to $\Gamma$. A similar situation was recently considered in~\cite{Carm11}, where steady-state inversion of a TLA in a cavity was predicted. In our scenario the cavity mode diverts a significant portion of the energy from the atom, which would otherwise contribute to the fluorescent light. This yields a reduction of the atomic excitation. From this point of view a not too good cavity is needed, in order to avoid a too strong backaction onto the atom. On the other hand, it has to be good enough to preserve the coherence of the atom, which would be lost in free-space  fluorescence. 

The coherent part of the atomic excitation (CPAE), $|\langle A_{12}\rangle|^2$, is determined by the interaction of the atom with both the cavity mode and the
vacuum modes in free space, leading to the fluorescence.
As a consequence of the above discussion, for the resonance condition~(\ref{eq.cavres}), the CPAE is increased due to the coupling to the cavity. Hence, we expect $\langle\hat A_{22}\rangle$ to decrease, while $|\langle\hat A_{12}\rangle|^2$ increases, yielding an obvious purification of the atomic state, according to Eqs.~(\ref{eq.CSI}),~(\ref{eq.purity}). A critical condition in this setup is the requirement of $\kappa\gg\Gamma$ and $g\approx\kappa$, so that the atom-cavity coupling significantly exceeds the atomic decay, $g\gg\Gamma$.
Experimental works, such as~\cite{Hood}, suggest that large ratios of $g/\Gamma$ can be achieved in the optical frequency range. In this experiment a value of  $g/\Gamma\approx23$ was realized, which will be used in our calculations reported below.

Combining the above arguments, we can determine the regime of optimized squeezing in cavity-assisted resonance fluorescence. We need strong pumping and atomic detuning, such that the atomic excitation in free space 
would slightly exceed $1/4$. For example, this would be the case for $|\delta_\text a|\gtrapprox|\Omega_\text R|$ and $\langle\hat A_{22}\rangle\lessapprox1/3$,
for details see~\cite{WelVo}. 
The cavity, which obeys the condition $\kappa\gg \Gamma$, is tuned to a resonance according to Eq.~(\ref{eq.cavres}). This yields a reduction of the atomic excitation to approximately $1/4$, as needed for 
maximal squeezing according to Eq.~(\ref{abs}).
The CPAE increases, resulting in a purification of the atomic state together with optimized squeezing, cf. Eq.~(\ref{eq.Sq-Pur}). The cavity emission rate $\kappa$ is chosen to be intermediate, $\kappa\approx g$, since for large $\kappa$~values 
the situation becomes close to that in free space. For smaller cavity 
losses the atomic excitation is not sufficiently reduced, due to the backaction of the cavity photons on the atom.

\begin{figure}[h]
  \includegraphics[width=8cm]{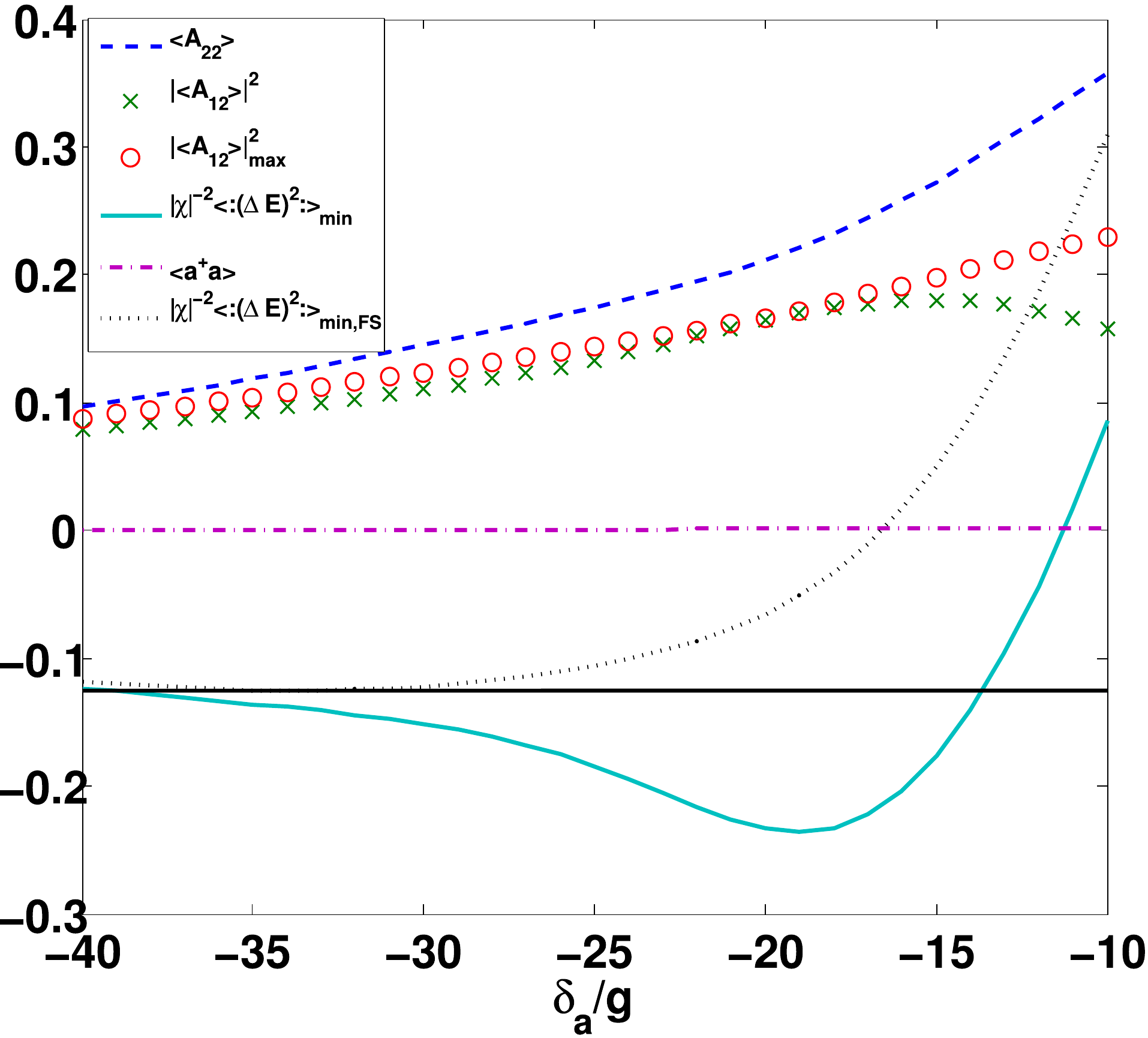}
  \caption{(color online). Dependence of the following quantities on $\delta_\text a$: atomic excitation $\langle\hat A_{22}\rangle$ (blue, dashed line),  CPAE $|\langle\hat A_{12}\rangle|^2$ (green crosses) and maximum CPAE $\langle\hat A_{11}\rangle\langle\hat A_{22}\rangle$ (red circles), cavity excitation $\langle\hat a^\dagger\hat a\rangle$ (violet, dashed-dotted line), normally ordered field variance of the fluorescence of a TLA in the cavity (light blue, solid curve), and in free space (black, dotted curve). The straight line (black, solid) marks the free space maximal squeezing of $-1/8$. The system parameters are: $\Omega_{\text R}/g=14$, $\kappa/g=1.58$, $\Gamma/g=0.04\overline{3}$, $\delta_\text c/g=-34$.}\label{fig.Sqmax}
\end{figure}

\paragraph*{Numerical results.}
In Fig.~\ref{fig.Sqmax} we show the dependence of the  atomic excitation and CPAE on the atomic detuning $\delta_{\rm a}$. The other parameters are chosen according to the requirements discussed above. The atomic excitation $\langle \hat A_{22} \rangle$ behaves similar to the free-space scenario. For a wide parameter range, the CPAE is rather close to its maximum value, cf. Eq.~(\ref{eq.CSI}), indicating significant purity of the atomic state. The purity is given by $\text{Tr}\{\hat\sigma^2\}{=}1{-}2(|\langle\hat A_{12}\rangle|^2_\text{max}{-}|\langle\hat A_{12}\rangle|^2)$. The cavity excitation is very small, even at the cavity resonance according to Eq.~(\ref{eq.cavres}),
occurring for the chosen parameters at $\delta_\text a\approx-19g$. This is consistent with the results of~\cite{FrQu} at the cavity resonance, which would predict $\langle\hat a^\dagger\hat a\rangle< 10^{-2}$ for our parameters. 

Around the cavity resonance we see, that the normally ordered field variance attains
the value of $-0.236$. This is more than 94\% of the maximum possible squeezing
of $-1/4$, which can be readily measured as discussed in~\cite{Supp}.
The purity Tr$\{\hat\sigma^2\}$ of the atomic subsystem (not depicted in Fig.~\ref{fig.Sqmax}), shows a clear maximum of about 99.5\% at the cavity resonance. The corresponding atomic excitation is $\langle\hat A_{22}\rangle\approx0.220$. 
Our numerical results for the obtained squeezing effect, the excitation and purity of the atomic state are in full agreement with the analytical relations of these quantities as given
in Eq.~(\ref{eq.Sq-Pur}) together with (\ref{var-min}).
For larger values of $g/\Gamma$, even more than 99\% of the absolute squeezing limit, $\langle:(\Delta\hat E)^2:\rangle_{\text{abs}}$,  can be achieved. This exceeds the free-space result substantially.

Let us now consider, for our optimized environment, the sensitivity of squeezing with respect to dephasing, which is crucial in free-space fluorescence. For this purpose we assume that, in addition to dephasing due to radiative damping, 
there is also radiationless dephasing described by the rate $\Gamma_\text D$.
For our considerations of the cavity-assisted atomic fluorescence we supplement 
the equations of motion~(\ref{eqmo}) with another Lindblad-term,
\begin{equation}
     \frac{d\hat\rho}{dt}=\frac{1}{i\hbar}[\hat H,\hat\rho]+\frac{\Gamma}{2}\mathcal{L}_{\hat A_{12}}[\hat\rho]+\frac{\Gamma_\text D}{2}\mathcal{L}_{\hat A_{22}}[\hat \rho]+\frac{\kappa}{2}\mathcal{L}_{\hat a}[\hat\rho].
\end{equation}
The additional dephasing only increases the decay of the off-diagonal matrix elements of the density operator, that is, the atomic coherence and hence the CPAE.

In free space, squeezing does not occur anymore if the dephasing rate $\Gamma_\text D$ exceeds the energy relaxation $\Gamma$. The atomic coherence needed for squeezing decays on a time scale which is faster than that of the emission of the fluorescence radiation. As stated above, in our setup, on a cavity resonance the CPAE is increased due to the increased transition amplitudes between atom and cavity. The backaction of the cavity onto the atom preserves the coherence on a longer time scale, thus increasing the actual coherence in the steady state. Hence, the robustness against dephasing is enhanced. In this context it is important that the cavity emission rate $\kappa$ significantly exceeds the atomic decay rate $\Gamma$. This renders it possible to observe squeezing in cavity-assisted resonance fluorescence for surprisingly strong atomic dephasing.

\begin{figure}[h]
  \includegraphics[width=8cm]{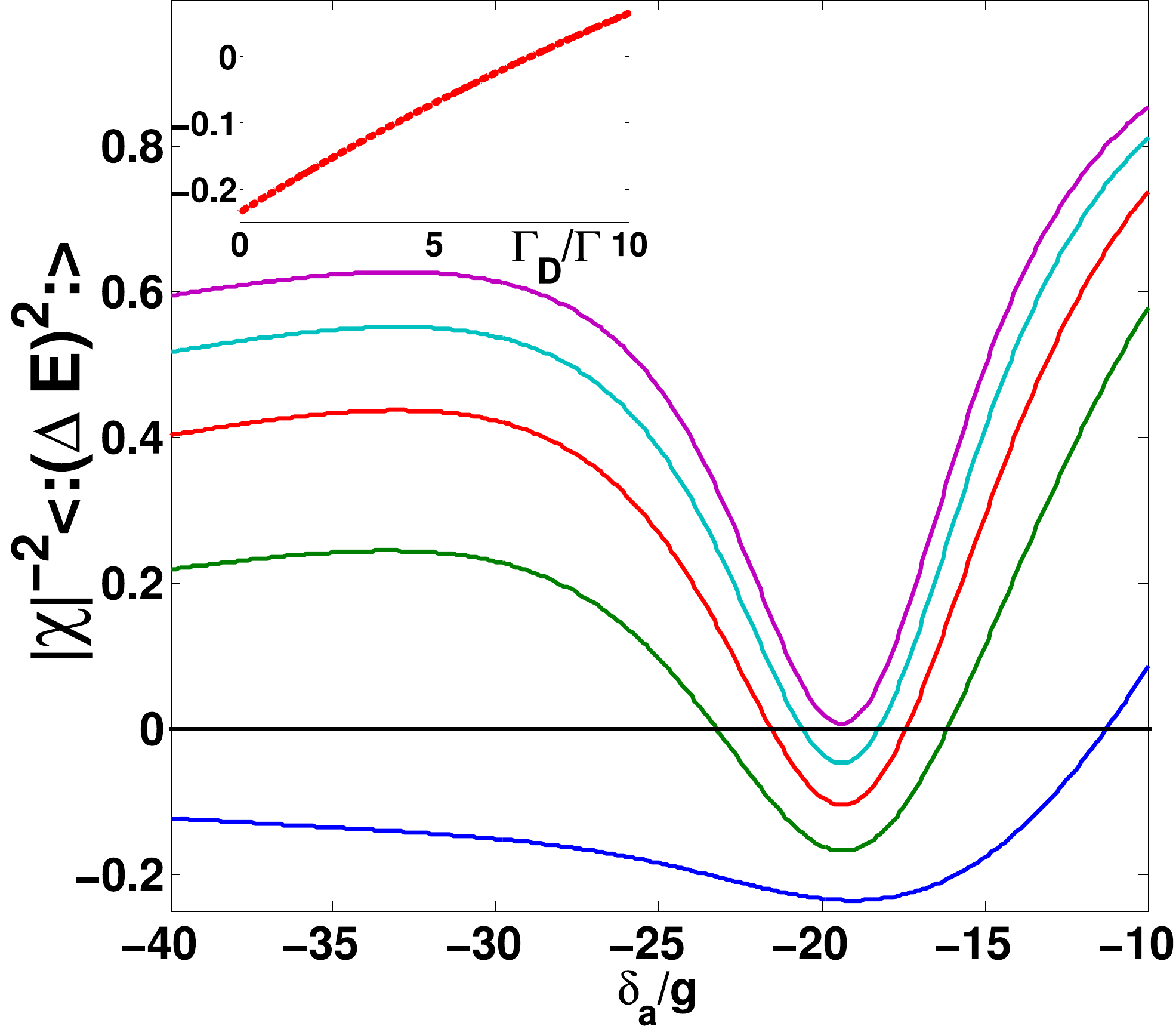}
  \caption{(color online). Normally ordered field variance over $\delta_a$ for different dephasings $\Gamma_\text D$. From bottom to top: $\Gamma_\text D/\Gamma=0,2,4,6,8$. The inset shows the dependence of the squeezing on $\Gamma_\text D$ at the cavity resonance, $\delta_a=-19g$. All other parameters are as in Fig.~\ref{fig.Sqmax}.}\label{fig.Deph}
\end{figure}

In Fig.~\ref{fig.Deph}, we show the normally ordered field variance for the same cavity as in Fig.~\ref{fig.Sqmax}, for different values of $\Gamma_\text D$. The dependence of the minimal field variance on $\Gamma_\text D$ is shown in the inset.
For $\Gamma_\text D<3.24\Gamma$, the minimal variance is below $-1/8$. This is the maximal squeezing in free space, which could only be achieved for $\Gamma_\text D=0$. The squeezing in the cavity setup under study vanishes for $\Gamma_\text D\approx7.47\Gamma$. This result may be of great importance for applications in light-emitting systems other than single atoms.
For example, significant dephasing is usually expected to occur in condensed matter systems. A typical candidate could be quantum dots in semiconductor microstructures. The scenario under study opens the possibility to create squeezing for relatively strong dephasing and intermediate cavity coupling, as it is typical for semiconductor microcavities~\cite{Khitrova,Finley,Michler,Dousse}.

\paragraph*{Conclusions.} We have studied the possibility of the optimization of squeezing  in the resonance fluorescence of a two-level atom. It is based on 
a special design of the environment of the atom, which can be achieved by a cavity with properly adjusted parameters.  The maximal squeezing of the light emitted from a coherently driven atom in an optimal environment is twice as strong as the maximal squeezing in free space experiments. More importantly, squeezing can be optimized for different atomic excitations, so that it is no longer limited to weakly 
driven systems. The resulting possibility to increase both the fluorescence intensity and the squeezing will substantially widen the possibility of detection and application of squeezing in resonance fluorescence. The realization of optimal squeezing is accompanied by the purification 
of the atomic state for different atomic excitations, which can be relevant for any other application which requires pure atomic states for different mean atomic excitations. It is also important that the strong limitation of squeezing by dephasing can be substantially reduced in our system. This may open important possibilities to control the emission of squeezed light from more complex,
miniaturized systems, such as quantum dots in semiconductor microcavities. 

\paragraph*{Acknowledgments.} The authors gratefully acknowledge stimulating conversation with P. R. Rice and J. Sperling. This work was supported by the Deutsche Forschungsgemeinschaft through SFB~652.

\section*{Supplemental Information}
\paragraph*{Numerical calculations.}
Starting from the master equations, we can formulate equations of motion for the elements of the density matrix 
\begin{equation}
  \varrho_{n,i;m,j}=\langle n,i|\hat \varrho|m,j\rangle,
\end{equation}
with the first index being the cavity photon number and the second being the atomic excitation ($i,j=1,2$). The explicit equations can be written as
\begin{widetext}
\begin{align}
 \dot\varrho_{n,1;m,1}=&-[i\delta_\text c(n-m)+\tfrac{\kappa}{2}(n+m)]\varrho_{n,1;m,1}-ig[\sqrt{n}\varrho_{n-1,2;m,1}-\sqrt{m}\varrho_{n,1;m-1,2}]\nonumber\\
		     &-i\Omega_\text R[\varrho_{n,2;m,1}-\varrho_{n,1;m,2}]+\Gamma\varrho_{n,2;m,2}+\kappa\sqrt{(n+1)(m+1)}\varrho_{n+1,1;m+1,1},\\
\dot\varrho_{n,1;m,2}=&[i(\delta_\text a-(n-m)\delta_\text c)-\tfrac{\Gamma+\kappa(n+m)}{2}]\varrho_{n,1;m,2}-ig[\sqrt{n}\varrho_{n-1,2;m,2} -\sqrt{m+1}\varrho_{n,1;m+1,1}]\nonumber\\
		      &- i\Omega_\text R(\varrho_{n,2;m,2}-\varrho_{n,1;m,1})+\kappa\sqrt{(n+1)(m+1)}\varrho_{n+1,1;m+1,2},\\
  \dot\varrho_{n,2;m,1}=&-[i(\delta_\text a+(n-m)\delta_\text c)+\tfrac{\Gamma+\kappa(n+m)}{2}]\varrho_{n,2;m,1}-ig[\sqrt{n+1}\varrho_{n+1,1;m,1} -\sqrt{m}\varrho_{n,2;m-1,2}]\nonumber\\
		      &- i\Omega_\text R(\varrho_{n,1;m,1}-\varrho_{n,2;m,2})+\kappa\sqrt{(n+1)(m+1)}\varrho_{n+1,2;m+1,1},\\
  \dot\varrho_{n,2;m,2}=&-[i\delta_\text c(n-m)+\Gamma+\tfrac{\kappa}{2}(n+m)]\varrho_{n,2;m,2}-ig[\sqrt{n+1}\varrho_{n+1,1;m,2}-\sqrt{m+1}\varrho_{n,2;m+1,1}]\nonumber\\
		      &-i\Omega_\text R(\varrho_{n,1;m,2}-\varrho_{n,2;m,1})+\kappa\sqrt{(n+1)(m+1)}\varrho_{n+1,2;m+1,2}.
\end{align}
\end{widetext}
We truncate the set of equations at a sufficiently large photon number $N$. By varying $N$, the validity of the calculations can be checked. Using ${\rm Tr} \hat \varrho =1$, we can eliminate one element of the main diagonal, in our case, we chose $\varrho_{0,1;0,1}$.  This introduces an inhomogeneity into the equations, allowing us to calculate the steady state density matrix simply by inverting the matrix of coefficients and multiplying with the inhomogeneity. Finally, the expectation values of interest can be directly obtained.

\paragraph*{Measurement of Squeezing.} The normally ordered variance of a light field is usually measured via  balanced homodyne detection, see e.g.~\cite{WelVo}. In case of single atom fluorescence the complications stem from the small collection efficiency, substantially reducing the effect to be measured. This problem can be resolved by correlation measurements~\cite{vogel,vogel95}, since in this case all correlation functions include the quantum efficiencies only as proportionality factors. One may detect the intensity correlation function
\begin{equation}
  \mathcal G^{(2,2)}(t_1,t_2)=\langle\hat E_1^{(-)}(t_1)\hat E_2^{(-)}(t_2)\hat E_2^{(+)}(t_2)\hat E_1^{(+)}(t_1)\rangle
\end{equation}
by the scheme in Fig.~2 of Ref.~\cite{vogel95}.
Here the indices $1,2$ refer to the different output channels of a beamsplitter. Following~\cite{vogel95}, we consider the difference between correlations at equal times and the corresponding steady state value,
\begin{equation}
  \Delta\mathcal G^{(2,2)}(t)=\mathcal G^{(2,2)}(t,t)-\lim_{\tau\rightarrow\infty}\mathcal G^{(2,2)}(t,t+\tau).
\end{equation}
For (stationary) single atom fluorescence as the signal and a 50/50 beamsplitter we obtain
\begin{equation}
  \Delta\mathcal G^{(2,2)}=-\frac{1}{4}\left(I_{\text{fl}}^2+I_{\text{lo}}\langle:(\Delta\hat E_{\text{fl}})^2:\rangle\right),
\end{equation}
with 'fl' and 'lo' denoting fluorescence and local oscillator, respectively, and $I$ is the intensity. For $I_\text{lo}\gg I_\text{fl}$, the second term is dominant. Squeezing is detected if $\Delta\mathcal G^{(2,2)}>0$.  The effect of our optimization of squeezing, which yields a factor of two, is clearly observed by this method.

\end{document}